\documentclass[aps,prl,twocolumn,superscriptaddress,nofootinbib,floatfix]{revtex4-2}

\usepackage{amsmath,amssymb}

\usepackage{braket}
\usepackage{graphicx}
\usepackage{xcolor}
\usepackage[
  colorlinks=true,
  linkcolor=blue,
  citecolor=blue,
  urlcolor=blue
]{hyperref}
\newcommand{\ii}{\mathrm{i}}
\newcommand{\dd}{\mathrm{d}}
\newcommand{\bl}{\boldsymbol{\lambda}}

\begin{document}
\title{Anomalous entanglement scaling from eigenvector nonorthogonality in critical non-Hermitian free fermions}

\author{Zhenyu Xiao}
\email{zyxiao@princeton.edu}
\affiliation{Princeton Quantum Initiative, Princeton University, Princeton, New Jersey 08544, USA}

\author{Shinsei Ryu}
\affiliation{Department of Physics, Princeton University, Princeton, New Jersey 08544, USA}

\date{\today}

\begin{abstract}
Entanglement carries universal content that labels phases and critical points.
We study the entanglement entropy of the steady states of critical non-Hermitian free-fermion chains. 
It scales logarithmically with subsystem size, but the coefficients vary continuously with the parameters and form a R\'enyi family that no single central charge can reproduce.
We trace this anomaly to an ``imaginary'' Dirac point, a crossing in the imaginary part of the energy where the occupied state switches between two Bloch states.
Their nonorthogonality weakens the occupation discontinuity and lowers the logarithmic coefficient.
A low-energy expansion yields closed-form coefficients in excellent agreement with lattice numerics in various one-dimensional critical steady states.
Remarkably, weak real onsite disorder leaves this logarithmic scaling intact and enhances the entanglement.
Our results provide a generic understanding of entanglement in critical non-Hermitian free-fermion steady states.
\end{abstract}

\maketitle

{\it Introduction.---}
Entanglement is a central window onto correlated quantum matter. 
Entanglement entropy often carries universal content that labels phases and critical points~\cite{Amico2008,Eisert2010}.
The idea is precisely realized at one-dimensional (1D) quantum critical points. 
In a critical ground state, the von Neumann entanglement entropy of a subsystem of length $\ell$ grows logarithmically as $S_1\sim(c/3)\ln\ell$ under periodic boundary conditions (PBCs), with a coefficient fixed by the central charge $c$ of the underlying conformal field theory (CFT)~\cite{Holzhey1994,Calabrese2004,Vidal2003}.

This diagnostic reaches well beyond ground states.
Entanglement entropy also characterizes unitary quantum dynamics, including thermalization~\cite{Calabrese2005,Deutsch1991,Srednicki1994,Rigol2008}, many-body localization~\cite{Bardarson2012,Nandkishore2015,Abanin2019}, and dynamical phase transitions~\cite{Heyl2013,Heyl2018}.
However, many systems probed in current experiments evolve non-unitarily~\cite{Diehl2008,Mueller2012,Daley2014}.
The effective dynamics is generated by non-Hermitian Hamiltonians, which arise in post-selected measurement trajectories~\cite{Dalibard1992,Plenio1998,Daley2014}, gain--loss processes in photonic and acoustic platforms~\cite{Ruter2010,Regensburger2012,Feng2017,ElGanainy2018,Miri2019,Ozawa2019}, and reduced descriptions of open quantum systems~\cite{Gorini1976,Lindblad1976,Plenio1998,Breuer2002,Daley2014,Ashida2020}.
Such Hamiltonians host phenomena with no Hermitian counterpart, including exceptional points~\cite{Heiss2012,Miri2019,Bergholtz2021}, parity--time symmetry breaking~\cite{Bender1998,Bender2007,Ruter2010,Regensburger2012,ElGanainy2018}, and the non-Hermitian skin effect~\cite{Yao2018,Kunst2018,Kawabata2019,Yokomizo2019,Okuma2020,Bergholtz2021}.

While the subject remains far less characterized than its Hermitian counterpart, existing studies of non-Hermitian entanglement span diverse definitions and dynamical settings~\cite{Herviou2019,Chang2020,Chen2021,Guo2021,Okuma2021,Modak2021,Bacsi2021,Gopalakrishnan2021,Jian2021YangLee,Lee2022,Tu2022,Ortega2022,Cipolloni2023,Hsieh2023,Kawabata2023,LeGal2023,TurkeshiSchiro2023,Zerba2023,Fossati2023,Rottoli2024,Feng2023SkinEffect,MunozArboleda2024,Yang2024,Chen2024,Zhou2024Quasicrystal,Zhou2024Floquet,Zhou2024Kitaev,Huang2025Feedback,Liu2025}.
In hybrid quantum circuits and continuously monitored systems~\cite{XiaoOhtsukiKawabata2025}, measurement-induced entanglement transitions whose critical scaling is logarithmic have been found, governed by boundary or non-unitary CFTs~\cite{Li2018,Skinner2019,Chan2019,Li2019,Bao2020,Choi2020,Jian2020,Gullans2020,Alberton2021,Fuji2020,Buchhold2021,Lavasani2021}.
Biorthogonal entanglement entropy can scale logarithmically in certain non-Hermitian ground states, with coefficients set by negative central charges of non-unitary CFTs~\cite{Brody2014,Chang2020,Bianchini2015a,Bianchini2015b,Bianchini2016,Couvreur2017,Tu2022,Hsieh2023}.
Parameter-dependent logarithmic coefficients of the von Neumann entropy $S_1$ have been found in non-Hermitian free-fermion and spin models~\cite{Turkeshi2021,TurkeshiSchiro2023,Zhou2024Kitaev}.
By contrast, a generic mechanism and classification for logarithmic entanglement in critical non-Hermitian free-fermion steady states remain lacking.


In this Letter, we study the entanglement entropy of the steady states of critical 1D non-Hermitian free-fermion chains, the right eigenstates selected at long times by the non-unitary evolution.
We find logarithmic scaling, $S_n=\zeta_n\ln\ell+O(1)$ ($n$ the R\'enyi index), and $\zeta_n$ drifts continuously below the $c=1$ value and forms a family that no single central charge can reproduce.
We trace this anomaly to an ``imaginary'' Dirac point, a band crossing in $\operatorname{Im}E(k)$, where the nonorthogonality of the right eigenstates weakens the occupation jump and yields a closed-form $\zeta_n$ in quantitative agreement with lattice numerics.
We further demonstrate that this scaling is model-independent: distinct band crossings contribute additively, generalizing to any 1D critical non-Hermitian steady state.
We also find that weak real onsite disorder leaves the logarithmic scaling intact and enhances the entanglement, unlike strong localization in Hermitian disordered free fermions~\cite{Anderson1958}.

\begin{figure}[thpb]
    \centering
    \includegraphics[width=1\columnwidth]{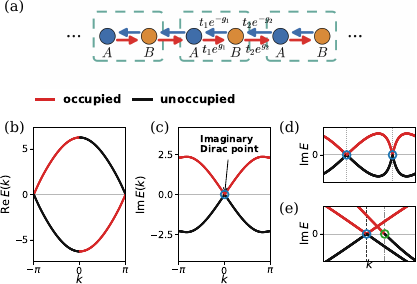}
    \caption{(a) Schematic of the non-Hermitian SSH model [Eq.~\eqref{eq:Hreal}].
    (b,c) Real and imaginary parts of the complex band structure $E_\pm(k)$ for $(t_1,t_2,g_1,g_2)=(3.6,2,1,0)$.
    In (c), the steady state fills the modes with $\operatorname{Im}E(k)>0$, and
    the band crossing at $\operatorname{Im}E_\pm(k) = 0$ defines the imaginary Dirac point.
    (d) A schematic of two imaginary Dirac points. (e) One imaginary Dirac point and one single-band Fermi point.
    }
    \label{fig:model-spectrum}
\end{figure}

\emph{Models and steady states.---}
We first consider a 1D non-Hermitian Su--Schrieffer--Heeger (SSH) model [Fig.~\ref{fig:model-spectrum}\,(a)]~\cite{Su1979,Esaki2011,Schomerus2013,Zhu2014,Lee2016,Klett2017,Weimann2017,Lieu2018,Yin2018,MartinezAlvarez2018,Shen2018,Gong2018,Klett2018,Parto2018,Kunst2018,Yao2018,Xiong2018,Kunst2019,Yokomizo2019,Okuma2020,Ghatak2020,Helbig2020,Bergholtz2021} with Hamiltonian
\begin{align} \label{eq:Hreal}
\hat{H}=\sum_j\Big[
&t_1e^{g_1}c_{j,b}^\dagger c_{j,a}
+t_1e^{-g_1}c_{j,a}^\dagger c_{j,b}
\nonumber\\
&+t_2e^{g_2}c_{j+1,a}^\dagger c_{j,b}
+t_2e^{-g_2}c_{j,b}^\dagger c_{j+1,a}
\Big] ,
\end{align}
where $c_{j,a}$ and $c_{j,b}$ annihilate spinless fermions on the two sublattices of unit cell $j$, satisfying canonical anticommutation relations, and $t_1,t_2,g_1,g_2\in\mathbb{R}$.
The imaginary gauge factors $e^{\pm g_\alpha}$ ($\alpha=1,2$) make the forward and backward hoppings asymmetric and render $\hat{H}$ non-Hermitian; at $g_1=g_2=0$ one recovers the standard Hermitian SSH chain.
We work under PBCs, setting aside the non-Hermitian skin effect (the accumulation of extensive eigenstates at one edge under open boundaries)~\cite{Yao2018}.

Long-time non-Hermitian evolution selects a steady state $\ket{\Psi_S}\propto\lim_{t\to\infty}e^{-\ii\hat{H}t}\ket{\Psi_0}$.
Decomposing the initial state in the right eigenbasis $\ket{R_\alpha}$ defined by $\hat{H}\ket{R_\alpha}=E_\alpha\ket{R_\alpha}$ with $E_\alpha\in\mathbb{C}$ and $\braket{R_\alpha|R_\alpha}=1$, each component grows as $e^{(\operatorname{Im}E_\alpha)t}$, so the one with the largest $\operatorname{Im}E_\alpha$ dominates at late times.
For a generic $\ket{\Psi_0}$ not restricted to a particular symmetry sector, and assuming the maximum imaginary part is unique,
\begin{equation}
\ket{\Psi_S}=\ket{R_{\alpha_\star}},
\qquad
\alpha_\star=\arg\max_\alpha\operatorname{Im}E_\alpha .
\label{eq:steady-state}
\end{equation}
The entanglement entropy of a subregion $A$ in the steady-state density matrix $\rho=\ket{\Psi_S}\bra{\Psi_S}$ is $S_n=(1-n)^{-1}\ln\operatorname{Tr}\rho_A^n$ with $\rho_A=\operatorname{Tr}_{\bar A}\rho$, where $n$ is the R\'enyi index and the von Neumann entropy is recovered as $n\to1$.
This convention differs from the biorthogonal one, in which the density matrix combines right and left eigenstates of $\hat{H}$ and is in general non-Hermitian; here, we keep $\rho=\ket{\Psi_S}\bra{\Psi_S}$ to access the physical entanglement of the time-evolved state.

\emph{Anomalous entanglement scaling.---}
The Hamiltonian $\hat{H}$ in Eq.~\eqref{eq:Hreal} is quadratic, $\hat{H}=\sum_{ij}H_{ij}c_i^\dagger c_j$.
If $H\phi_m^R=\varepsilon_m\phi_m^R$, then $d_m^\dagger=\sum_i(\phi_m^R)_i c_i^\dagger$ creates a single-particle right eigenmode and the many-body right eigenstates are Slater determinants $\prod_{m\in S}d_m^\dagger\ket{0}$.
Equation~\eqref{eq:steady-state} therefore identifies the steady state with the Fock configuration~\cite{XiaoKawabata2026}
\begin{equation}
S_\star=\{m:\operatorname{Im}\varepsilon_m>0\}.
\label{eq:positive-gain-configuration}
\end{equation}
Under PBCs, these modes follow from the Bloch Hamiltonian
\begin{equation}
H(k)=
\begin{pmatrix}
0&h_+(k)\\ h_-(k)&0
\end{pmatrix},
\quad
\begin{aligned}
h_+(k)&=t_1e^{-g_1}+t_2e^{g_2}e^{-\ii k},\\
h_-(k)&=t_1e^{g_1}+t_2e^{-g_2}e^{\ii k} .
\end{aligned}
\label{eq:bloch}
\end{equation}
A sublattice symmetry~\cite{Kawabata2019} $H(k)=-\sigma_z H(k)\sigma_z$ pairs the bands as $E_\pm(k)=\pm\sqrt{h_+(k)h_-(k)}$.

We call $\hat{H}$ gapless when $\operatorname{Im}E(k)=0$ at isolated momenta, the non-Hermitian analog of Fermi points, and focus on this case; a twisted boundary condition keeps every allowed momentum off the zeros, so Eq.~\eqref{eq:positive-gain-configuration} fixes $\ket{\Psi_S}$ uniquely by filling all modes with $\operatorname{Im}E(k)>0$ [Fig.~\ref{fig:model-spectrum}(b)].
We set aside the gapped case (area-law entanglement) and the $\mathcal{PT}$-symmetric regime~\cite{Bender1998,Bender2007,ElGanainy2018}, in which $\operatorname{Im}E(k)$ vanishes over a finite $k$-range and $\ket{\Psi_S}$ is not unique.

The entanglement entropy of a fermion Slater-determinant state is fully captured by the ordinary correlation matrix $C_{i\alpha;j\beta}=\bra{\Psi_S}c_{j\beta}^\dagger c_{i\alpha}\ket{\Psi_S}$.
In the present two-band problem, each momentum carries a single occupied right Bloch spinor $u^R(k)$, selected by $\operatorname{Im}E(k)>0$ and normalized as $(u^R)^\dagger u^R=1$, so on a ring of $L$ unit cells the real-space correlator is the Fourier transform of the occupied projector,
\begin{equation}
C_{i\alpha;j\beta}=\frac{1}{L}\sum_k e^{\ii k(i-j)} u^R_\alpha(k)\,[u^R_\beta(k)]^* .
\label{eq:bloch-correlation}
\end{equation}
In more general scenarios, the occupied right eigenstates are nonorthogonal, and $C$ must be formed after orthonormalizing the occupied subspace, as described in the End Matter.
For a contiguous interval $A$ of $\ell$ unit cells, the standard correlation-matrix formula~\cite{Peschel2003,Peschel2009} relates the R\'enyi entropies to the eigenvalues $\nu_a$ of the restricted correlation matrix $C_A$,
\begin{equation}
S_n(A)=\sum_a h_n(\nu_a),\quad h_n(\nu)=\frac{1}{1-n}\ln[\nu^n+(1-\nu)^n].
\label{eq:renyi}
\end{equation}

\begin{figure}[t]
    \centering
    \includegraphics[width=0.9\columnwidth]{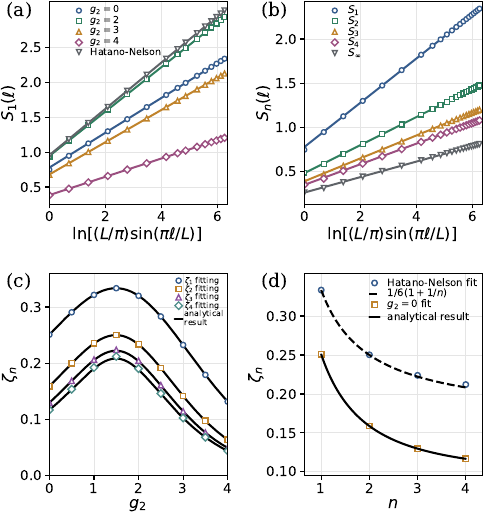}
    \caption{
    (a) Von Neumann entanglement entropy \(S_1\) versus chord distance \(d_\ell=(L/\pi)\sin(\pi\ell/L)\) at \((t_1,t_2,g_1)=(3.6,2,1)\) for several \(g_2\), and the Hatano--Nelson point \((2,2,1,1)\).
    (b) R\'enyi entropies \(S_n\) at \(g_2=0\); fitted \(\zeta_n\) tabulated in Eq.~\eqref{eq:zeta-comparison}.
    (c) \(\zeta_n\) versus \(g_2\) for \(n=1,2,3,4\). 
    (d) \(\zeta_n\) versus \(n\) at \(g_2=0\), compared with the Hatano--Nelson value \((1+1/n)/6\).
    Numerics use \(L=65536\) with linear fits over \(128\le\ell\le512\).
    In (c) and (d), dots come from numerical fitting and solid lines are the analytical prediction [Eqs.~\eqref{eq:bloch}, \eqref{eq:theta}, \eqref{eq:zeta-integral}].}
    \label{fig:clean-scaling}
\end{figure}

We numerically compute $S_n(\ell,\bl)$ of $\ket{\Psi_S}$ for different interval lengths $\ell$ and parameter sets $\bl=(t_1,t_2,g_1,g_2)$ throughout the gapless regime.
For the parameter sets we examine, the data confirm a critical scaling [Figs.~\ref{fig:clean-scaling}\,(a) and (b)]
\begin{equation}
S_n(\ell,\bl)=\zeta_n(\bl)\ln d_\ell+O(1) ,
\label{eq:scaling-numerical}
\end{equation}
where $d_\ell=(L/\pi)\sin(\pi\ell/L)$ is the chord distance appropriate to the ring geometry~\cite{Holzhey1994,Calabrese2004}.
At the Hatano--Nelson point $t_1=t_2$, $g_1=g_2$, where $\hat{H}$ reduces to a single-band non-reciprocal chain, the fits give $\zeta_n=(1+1/n)/6$ [Fig.~\ref{fig:clean-scaling}\,(d)], the prediction of a $c=1$ CFT.
This is natural. 
There, the steady state is the ground state of the Hermitian chain $\sum_i(-\ii c_i^\dagger c_{i+1}+\text{h.c.})$, a single-band Fermi sea with the standard $c=1$ universality.
For generic $\bl$, however, $\zeta_n(\bl)$ becomes nonuniversal: it drifts continuously with $\bl$ [see Fig.~\ref{fig:clean-scaling}\,(c)], and the family across $n$ cannot be assigned any single central charge [Fig.~\ref{fig:clean-scaling}\,(d)].
The non-Hermitian steady state therefore remains critical but is generically distinct from any equilibrium 1D critical state.

\emph{Imaginary Dirac point.---}
To resolve this anomaly, we derive $\zeta_n(\bl)$ analytically from the long-distance form of the correlator [Eq.~\eqref{eq:bloch-correlation}], which is controlled by the momenta around the discontinuity of the occupied branch, where $\operatorname{Im}E(k)$ crosses zero.
For the non-Hermitian SSH chain, this occurs at $k=0$, where $\operatorname{Im}E_\pm(0)=0$ while $\operatorname{Re}E_\pm(0)=\pm\sqrt{h_+(0)h_-(0)}\ne0$ [Fig.~\ref{fig:model-spectrum}(b)].
The two bands therefore touch in $\operatorname{Im}E(k)$ while remaining split in real energy (the imaginary Dirac point).

We choose the sublattice gauge such that $h_\pm(0)>0$.
In lowest-order perturbation theory in $k$, the occupied right spinor equals its value at the crossing,
\begin{equation}
u_\sigma^R=\frac{1}{\sqrt{h_+(0)+h_-(0)}}\begin{pmatrix}\sqrt{h_+(0)}\\ \sigma\sqrt{h_-(0)}\end{pmatrix},\qquad \sigma=\pm ,
\label{eq:right-spinors}
\end{equation}
and only the occupation switches: the $\sigma=+$ branch is filled for $k<0$ and the $\sigma=-$ branch for $k>0$.
For a Hermitian crossing, $u_+^R$ and $u_-^R$ would be orthogonal.
Here, they need not be, and their overlap defines a single angle
\begin{equation}
\sin\theta\equiv|(u_+^R)^\dagger u_-^R|=\left|\frac{h_+(0)-h_-(0)}{h_+(0)+h_-(0)}\right| ,\qquad \cos\theta\ge0 .
\label{eq:theta}
\end{equation}

Define the momentum-space correlator $C_{\alpha\beta}(k)\equiv\bra{\Psi_S}c_{k,\beta}^\dagger c_{k,\alpha}\ket{\Psi_S}=u^R_\alpha(k)\,[u^R_\beta(k)]^*$, the occupied projector at each $k$.
Fixing the phase of $u_-^R$ so that $(u_+^R)^\dagger u_-^R=\sin\theta$, we choose an orthonormal basis in which $u_\pm^R=(\sqrt{(1+\sin\theta)/2},\,\mp\sqrt{(1-\sin\theta)/2})^{T}$, and the low-energy correlator becomes
\begin{equation} \label{eq:low-energy-correlator}
C(k)\simeq\tfrac12 I+\tfrac12\sin\theta\,\tau_z+\tfrac12\operatorname{sgn}(k)\cos\theta\,\tau_x ,
\end{equation}
with $\tau_\mu$ the Pauli matrices in this basis.
The discontinuity across the crossing is therefore $C(0^+)-C(0^-)=\cos\theta\,\tau_x$, with eigenvalues $\pm\cos\theta$.
In the Hermitian case, $\theta = 0$, the amplitude of the discontinuity is one; in generic non-Hermitian cases, the amplitude is reduced to $\cos\theta$ by the nonorthogonality of the eigenvectors.

The weakened discontinuity leads to reduced entanglement entropy.
Fourier transforming Eq.~\eqref{eq:low-energy-correlator} (the constant terms give $\delta_{ij}$, while $\operatorname{sgn}(k)$ gives the discrete kernel $\ii/[\pi(i-j)]$) yields the long-distance correlator
\begin{equation}
C_{ij}\simeq\left(\tfrac12 I+\tfrac12\sin\theta\,\tau_z\right)\delta_{ij}+\frac{\ii\cos\theta}{2\pi(i-j)}\,\tau_x ,
\label{eq:C0C1}
\end{equation}
where short-distance pieces shift only the nonuniversal $O(1)$ part of the entropy and are dropped.
On an interval of $\ell$ cells, the $\tau_x$ part of $C_{ij}$ becomes the kernel $K_{ij}=\ii/[2\pi(i-j)]$ and the $\tau_z$ part stays constant.
For large $\ell$, the eigenvalues of $2K_A$ are given asymptotically by $\eta_m=\tanh(\epsilon_m^{(0)}/2)$ with $\epsilon_m^{(0)}\simeq\pi^2(2m+1)/\ln\ell$~\cite{Peschel2004,Peschel2009}; diagonalizing $K_A$ then block-diagonalizes $C_A$ into the $2\times2$ band-space matrices
\begin{equation}
\tfrac12 I+\tfrac12\sin\theta\,\tau_z+\tfrac12\eta_m\cos\theta\,\tau_x ,
\label{eq:band-block}
\end{equation}
with eigenvalues $\nu_{m,\pm}=\tfrac12\pm\tfrac12\sqrt{\sin^2\theta+\cos^2\theta\,\eta_m^2}$.
The corresponding entanglement energies, defined by $\nu_{m,\pm}=(1+e^{-\epsilon_{m,\pm}})^{-1}$, obey
\begin{equation}
\tanh(\epsilon_{m,\pm}/2)=\pm\sqrt{\sin^2\theta+\cos^2\theta\,\tanh^2(\epsilon_m^{(0)}/2)} ,
\label{eq:spectrum-map}
\end{equation}
which is bounded away from zero: unlike the gapless Hermitian case ($\theta=0$), the entanglement spectrum is gapped, $|\epsilon|\ge E_\theta=2\operatorname{arctanh}|\sin\theta|$.

Inserting these eigenvalues into Eq.~\eqref{eq:renyi} and approximating the levels $\epsilon_m^{(0)}$ by a constant density $\ln\ell/(2\pi^2)$, the $\ln\ell$ coefficient is
\begin{equation}
\zeta_n=\frac{2}{\pi^2}\int_{E_\theta}^{\infty}\dd\epsilon\,\frac{\cos\theta\,\tanh(\epsilon/2)}{\sqrt{\tanh^2(\epsilon/2)-\sin^2\theta}}\,h_n\!\left(\frac{1}{1+e^{-\epsilon}}\right) ,
\label{eq:zeta-integral}
\end{equation}
with $h_n$ the single-level R\'enyi entropy [Eq.~\eqref{eq:renyi}].
At $\theta=0$ the gap closes and the integral returns the $c=1$ value $\zeta_n=(1+1/n)/6$.
For generic $\theta$, it evaluates in closed form for integer $n$ and $n\to\infty$; the resulting expressions are collected in Eq.~\eqref{eq:zeta-closed}.
For $n=1$, $3\zeta_1$ reproduces the closed-form effective central charge derived in Ref.~\cite{TurkeshiSchiro2023} for the steady state of the monitored non-Hermitian spin chain, which realizes an imaginary Dirac point after a Jordan--Wigner transformation.

Because $\theta$ is fixed by the band structure alone [Eq.~\eqref{eq:theta}], the closed form $\zeta_n(\theta)$ carries no adjustable parameter.
It nonetheless tracks the fitted coefficients for every $n=1,2,3,4$ as $g_2$ sweeps the gapless regime [Fig.~\ref{fig:clean-scaling}(c)], and at fixed parameters reproduces the entire R\'enyi family out to $n\to\infty$ to three or four significant figures [Fig.~\ref{fig:clean-scaling}(d) and Eq.~\eqref{eq:zeta-comparison}], a family that no single central charge can fit.
We thus demonstrate that the weakened discontinuity at the imaginary Dirac point accounts for the full $\ln\ell$ coefficient, the neglected corrections affecting only the nonuniversal $O(1)$ term.

\emph{Generic critical steady states.---}
We promote the above analysis to generic 1D non-Hermitian critical steady states, characterized by $\operatorname{Im}E_a(k)=0$ (with $a$ indexing bands) at finitely many isolated momenta.
Each such crossing is classified by the number of bands touching $\operatorname{Im}E=0$ at the same $k$.
For a single band, $\operatorname{Im}E$ changes sign across the crossing and the occupied Bloch state switches between filled and empty; the contribution matches that of a chiral free fermion ($c=1/2$), analogous to a Hermitian Fermi point~\cite{Banerjee2026}.
For two bands, the crossing is the imaginary Dirac point analyzed above, contributing $\zeta_n(\theta_i)$ with overlap angle $\theta_i$ [Eq.~\eqref{eq:theta}].
Here $\theta_i$ ranges from the Hermitian limit of orthogonal spinors ($\theta_i=0$) to an exceptional point where they coalesce ($\theta_i=\pi/2$; see Ref.~\cite{supplemental}).
When several such crossings coexist at distinct momenta, the contributions add [see Figs.~\ref{fig:model-spectrum}(d) and (e) for a schematic], and the $\ln\ell$ coefficient takes the universal form
\begin{equation}
\zeta_n = \sum_{i=1}^{m} \zeta_n(\theta_i) + N_F\,\frac{1+1/n}{12} ,
\label{eq:additive-zeta}
\end{equation}
where $m$ counts the imaginary Dirac points (with overlap angles $\theta_i$) and $N_F$ counts the single-band Fermi points.
This generalizes the Hermitian critical free-fermion result $\zeta_n=(c/6)(1+1/n)$, recovered as $\theta_i\to 0$ with $c=m+N_F/2$.
Imaginary Dirac points are symmetry-protected whenever a band pairing $E_+(k)=-E_-(k)$ is enforced, as for sublattice symmetry $H(k)=-\sigma_z H(k)\sigma_z$.
Numerical verification of Eq.~\eqref{eq:additive-zeta} across families of non-Hermitian critical free-fermion models is presented in Supplemental Material~\cite{supplemental}.

For a particle-conserving $\hat{H}$, we may also fix the steady state to a definite particle-number sector.
This is equivalent to taking the global steady state of $\hat{H}+\ii\mu\hat{N}$, where $\hat{N}=\sum_i c_i^\dagger c_i$ and the imaginary chemical potential $\ii\mu$ ($\mu\in\mathbb{R}$) uniformly shifts $\operatorname{Im}E$ to fix the filling.
The $\ln\ell$ coefficient again follows from Eq.~\eqref{eq:additive-zeta}~\cite{supplemental}.
The same formula applies to the non-Hermitian ground state~\cite{Herviou2019,Chang2020,Guo2021}, defined as the many-body eigenstate with the smallest $\operatorname{Re}E$, where we classify the crossings of $\operatorname{Re}E(k)=0$ instead of $\operatorname{Im}E(k)=0$~\cite{supplemental}.

\begin{figure}[hbt]
    \centering
    \includegraphics[width=0.9\columnwidth]{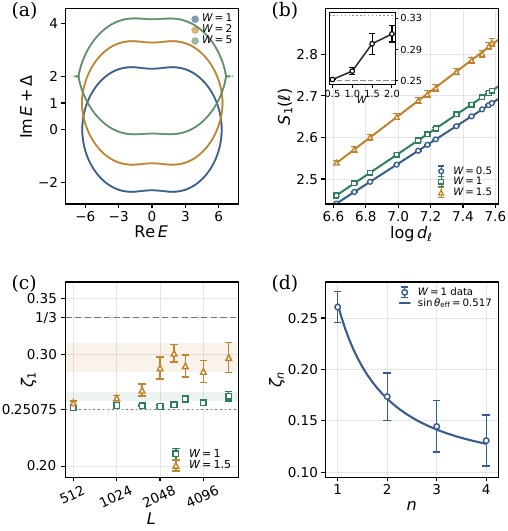}
    \caption{Real onsite disorder $W_{j,\mu}\in[-W/2,W/2]$ in the non-Hermitian SSH steady state.
    (a) Single-realization complex spectra at $L=512$ for $W=1,2,5$, shifted by $\Delta=0,1,2$ for clarity.
    (b) Disorder- and origin-averaged $S_1(\ell)$ versus $\ln d_\ell$ at $L=6144$, with the chord distance $d_\ell=(L/\pi)\sin(\pi\ell/L)$; lines are weighted linear fits, and the inset shows the fitted $\zeta_1$ versus $W$.
    (c) Finite-size convergence of $\zeta_1$ at $W=1$ and $W=1.5$; dotted and dashed lines mark the clean value $\zeta_1=0.25075$ and the $c=1$ Hatano--Nelson value $1/3$.
    (d) R\'enyi coefficients $\zeta_n$ at $W=1$ (symbols) compared with the single-parameter form Eq.~\eqref{eq:zeta-integral} at a fitted $\theta_{\rm eff}$ (line).}
    \label{fig:real-onsite-disorder}
\end{figure}

\emph{Disorder effect.---}
We add real onsite disorder
\begin{equation}
H_{\rm dis}=\sum_{j,\mu=a,b} W_{j,\mu}\,c_{j,\mu}^\dagger c_{j,\mu} ,
\end{equation}
with $W_{j,\mu}$ drawn independently from $[-W/2,W/2]$.
Following Eq.~\eqref{eq:positive-gain-configuration}, the occupied subspace is spanned by the right single-particle eigenstates of $\hat{H}+H_{\rm dis}$ with $\operatorname{Im}\varepsilon_m>0$.
These eigenstates are generically nonorthogonal, so building the right-state correlation matrix $C$ requires an additional orthonormalization step; we carry it out via the ordered Schur decomposition described in the End Matter.
The construction is well-defined as long as the complex spectrum stays separated from the $\operatorname{Im}E=0$ occupation boundary [Fig.~\ref{fig:real-onsite-disorder}(a)].
At larger $W$, real-eigenvalue ``wings''~\cite{Hatano1996} [Fig.~\ref{fig:real-onsite-disorder}(a)] develop, making the steady state nonunique. We therefore restrict the scaling analysis to $W\le 2$.

Figure~\ref{fig:real-onsite-disorder}(b) shows that, throughout this weak-disorder regime, the disorder-averaged $S_1(\ell)$ remains logarithmic in the chord distance, while $\zeta_1$ increases with $W$.
Finite-size analysis [Fig.~\ref{fig:real-onsite-disorder}(c)] shows that $\zeta_1$ at both $W=1$ and $W=1.5$ saturates for $L\gtrsim 3000$, with converged values lying between the clean value $\zeta_1(\theta)$ and the $c=1$ Hatano--Nelson value $1/3$.
The full R\'enyi family at $W=1$ is still described by Eq.~\eqref{eq:zeta-integral} after replacing $\theta$ by a fitted $\theta_{\rm eff}$ [Fig.~\ref{fig:real-onsite-disorder}(d)].
These results suggest that weak real onsite disorder renormalizes the occupation-discontinuity amplitude while preserving the logarithmic form.
This behavior contrasts with the standard 1D Hermitian free-fermion benchmarks: generic disorder gives Anderson-localized area-law states, while symmetry-tuned critical disorder flows to the infinite-randomness random-singlet fixed point with the universal coefficient $(\ln 2)/3$~\cite{Anderson1958,Fisher1994,Refael2004}.

Two complementary observations provide intuition for this robustness.
First, adding a uniform mass $m_0 I+m_z\sigma_z$ to the Bloch matrix at the Dirac point only shifts the real part, $E_\pm=m_0\pm\sqrt{h_+(0)h_-(0)+m_z^2}$, leaving $\operatorname{Im}E_\pm=0$ intact and preserving the crossing.
Second, a single impurity cannot resonantly backscatter a low-energy mode, as for a chiral fermion at a Hermitian Fermi point: generically each complex energy is attained by a single PBC state [Fig.~\ref{fig:real-onsite-disorder}(a)], leaving no distinct outgoing channel~\cite{Liu2021SingleImpurity,Leumer2025Impurity} (see the Supplemental Material for an explicit lattice scattering solution).

\emph{Discussion and outlook.---}
We have established a generic picture of entanglement in 1D critical non-Hermitian free-fermion steady states.
As in an equilibrium critical ground state, the logarithm comes from states near the occupation boundary $\operatorname{Im}E(k)=0$; but where two bands cross there, an imaginary Dirac point forms, and the nonorthogonality of the right eigenvectors weakens the projector discontinuity, opening an entanglement gap and a continuously tunable coefficient $\zeta_n(\theta)$ below the Hermitian Dirac value.
Multiple crossings contribute additively, so the formula covers generic critical steady states, and the scaling is robust to weak disorder, unlike its Hermitian counterpart.

The same coefficient arises in two very different settings: two partially transmitting defects bounding a subsystem of a critical free-fermion chain~\cite{Peschel2012,Calabrese2012}, and a single-shot weak measurement of the site occupations~\cite{Garratt2023,Sun2023Measurement,Weinstein2023MeasuredIsing,Yang2023Measurement}, both of which lower the entanglement while keeping it logarithmic.
There, $\cos\theta$ is replaced by the defect transmission or the measurement strength, and a boundary CFT describes each, with its boundary in space for the defect and in time for the measurement.
In each, the nonuniversal coefficient is a boundary effect: a localized perturbation dressing an otherwise standard $c=1$ bulk.
Our steady state hosts no such perturbation and is translation invariant in both space and time, yet it shares the same coefficient, making the anomaly a bulk effect.
Whether it admits an emergent boundary-CFT description remains open.

We have focused on PBCs because, under open boundaries, the non-Hermitian skin effect can substantially reshape steady states and their entanglement~\cite{Kawabata2019,Yokomizo2019,Okuma2020,Zhang2020,Okuma2021,Kawabata2023,Feng2023SkinEffect,Huang2025Feedback}.
This boundary sensitivity is separate from the imaginary-Dirac-point mechanism. 
The same logarithmic scaling occurs in open-boundary models as long as the skin effect is absent, as shown in Supplemental Material~\cite{supplemental}.

Whether the weakened-discontinuity picture survives interactions, for instance in non-Hermitian Luttinger liquids~\cite{Affleck2004}, is a further open question.
In higher dimensions the occupation boundary $\operatorname{Im}E(\mathbf{k})=0$ becomes a nodal hypersurface, and we expect a formula of the same Widom form~\cite{Widom1982,Gioev2006}; the subtle point is that the overlap angle is now a momentum-dependent $\theta(\mathbf{k})$ varying along the hypersurface.
A controlled effective theory for the disorder-renormalized $\theta_{\rm eff}$, its possible connection to a topological field theory, and the contrasting flow under complex onsite disorder (which appears to reach a different non-Hermitian fixed point) are natural directions for further study.
Experimentally, non-Hermitian time evolution can be realized through measurement and postselection in open quantum systems and quantum circuits~\cite{Ashida2020,Kawabata2023,Fleckenstein2022}.

{\it Acknowledgments.---}
Z.X. thanks Kohei Kawabata, Shaokai Jian, Ze Chen, and Shuo Liu for helpful discussions.
Z.X. is pleased to acknowledge that the work reported in this paper was substantially performed using the Princeton Research Computing resources at Princeton University.
Z.X. is supported by the Princeton Quantum Initiative Fellowship.
S.R. is supported by a Simons Investigator Grant from the Simons Foundation (Award No.\ 566116).
This work is supported by the Gordon and Betty Moore Foundation EPiQS initiative, Grant GBMF8685.01.

\bibliography{clean_limit_prl_refs}

\section*{End Matter}

\subsection{Closed-form coefficients and numerical comparison}
\label{sec:endmatter-zeta}
We collect the closed-form expressions for the logarithmic coefficients.
Similar integrals [Eq.~\eqref{eq:zeta-integral}] have been evaluated in Refs.~\cite{TurkeshiSchiro2023,Peschel2012,Calabrese2012}.
The resulting closed forms for the von Neumann coefficient and the R\'enyi coefficients for $n=2,3,\ldots,\infty$ are
\begin{widetext}
\begin{equation}
\begin{aligned}
\zeta_1&=\frac{4}{\pi^2}\left[\operatorname{Li}_2(1-\cos\theta)+\frac{\pi^2}{24}(3\cos\theta-1)+\frac{1+\cos\theta}{2}\left\{\operatorname{Li}_2[-\tan^2(\theta/2)]-\operatorname{Li}_2[\tan^2(\theta/2)]\right\}\right],\\
\zeta_n&=\frac{4}{\pi^2(n-1)}\sum_{r=1}^{\lfloor n/2\rfloor}\arcsin^2\!\left[\cos\theta\cos\frac{(2r-1)\pi}{2n}\right]\quad (n=2,3,\ldots),\\
\zeta_\infty(\theta)&=\frac{\operatorname{Li}_2(\cos^2\theta)}{\pi^2},
\end{aligned}
\label{eq:zeta-closed}
\end{equation}
\end{widetext}
where $\operatorname{Li}_2(x)=-\int_0^x \dd t\,\ln(1-t)/t$ is the dilogarithm.
These expressions recover the Hermitian free-fermion value at $\theta=0$:
$\zeta_n(0)=(1+1/n)/6$ and $\zeta_\infty(0)=1/6$.
At the opposite limit $\theta=\pi/2$, the two right eigenvectors coalesce, the projector discontinuity vanishes, and all logarithmic coefficients vanish.
For the SSH parameters $(t_1,t_2,g_1,g_2)=(3.6,2,1,0)$ used in the main text, these expressions give $\zeta_n^{\rm th}$, matching the lattice fits $\zeta_n^{\rm num}$ term by term:
\begin{equation}
\begin{array}{c|ccccc}
n&1&2&3&4&\infty\\
\hline
\zeta_n^{\rm th}& 0.25075 & 0.15878 & 0.12976 & 0.11672 & 0.08777\\
\zeta_n^{\rm num}& 0.25075 & 0.15880 & 0.12983 & 0.11682 & 0.08786 .
\end{array}
\label{eq:zeta-comparison}
\end{equation}
For other parameters (varying $g_2$), the analytical predictions [solid lines in Fig.~\ref{fig:clean-scaling}(c)] also match the lattice numerics [dots in Fig.~\ref{fig:clean-scaling}(c)] throughout the gapless regime.

\subsection{Schur decomposition for the disordered correlation matrix}
\label{sec:endmatter-schur}
For each disorder realization, we use the same occupation rule as in Eq.~\eqref{eq:positive-gain-configuration}.
Let $H_{\rm d}=H+H_{\rm dis}$ denote the disordered single-particle Hamiltonian.
If $H_{\rm d}\phi_m^R=\varepsilon_m\phi_m^R$, then $d_m^\dagger=\sum_i(\phi_m^R)_i c_i^\dagger$ creates a right single-particle mode.
Let $R_+$ be the matrix whose columns are the right eigenvectors with $\operatorname{Im}\varepsilon_m>0$.
These orbitals are generally not mutually orthogonal, so we orthonormalize the occupied subspace by a thin QR decomposition $R_+=Q_+\mathcal R$, where $Q_+^\dagger Q_+=I$ and $\mathcal R$ is upper triangular.
With $\tilde d_\alpha^\dagger=\sum_i(Q_+)_{i\alpha}c_i^\dagger$, the normalized Slater determinant built from $R_+$ is equivalently written as $\ket{\Psi_S}\propto\prod_\alpha\tilde d_\alpha^\dagger\ket{0}$, where $\alpha$ runs over the columns of $Q_+$~\cite{Cao2019,XiaoOhtsukiKawabata2025}.
Its correlation matrix is the occupied-subspace projector $C=Q_+Q_+^\dagger$.

To see how this is implemented numerically, suppose that the full set of right eigenvectors is ordered by descending $\operatorname{Im}\varepsilon_m$, so that
\begin{equation}
H_{\rm d}=RDR^{-1},\qquad D=\operatorname{diag}(\varepsilon_m).
\end{equation}
A QR decomposition of the ordered eigenvector matrix gives $R=Q\mathcal R$, where $Q$ is unitary and $\mathcal R$ is upper triangular.
Writing $Q=(Q_+,Q_-)$, the columns of $Q_+$ span the same occupied subspace obtained by QR-orthonormalizing the selected right eigenvectors.
Substituting $R=Q\mathcal R$ gives
\begin{equation}
H_{\rm d}=Q(\mathcal R D\mathcal R^{-1})Q^\dagger .
\end{equation}
Because $\mathcal R D\mathcal R^{-1}$ is upper triangular with the same diagonal entries as $D$, this is precisely the ordered Schur decomposition of $H_{\rm d}$, with $U=Q$ and $T=\mathcal R D\mathcal R^{-1}$.
Thus, in the numerics, we compute the ordered Schur decomposition directly and use its leading columns as $Q_+$.
\end{document}


\title{\textbf{\large Supplemental Material for ``Anomalous entanglement scaling from eigenvector nonorthogonality in critical non-Hermitian free fermions''}}

\author{Zhenyu Xiao}
\email{zyxiao@princeton.edu}
\affiliation{Princeton Quantum Initiative, Princeton University, Princeton, New Jersey 08544, USA}

\author{Shinsei Ryu}
\affiliation{Department of Physics, Princeton University, Princeton, New Jersey 08544, USA}

\date{\today}

\maketitle

\setcounter{secnumdepth}{3}
\renewcommand{\thesection}{S\Roman{section}}
\renewcommand{\thesubsection}{\thesection.\Alph{subsection}}
\renewcommand{\thesubsubsection}{\thesubsection.\arabic{subsubsection}}
\renewcommand{\theequation}{S\arabic{equation}}
\renewcommand{\thefigure}{S\arabic{figure}}
\renewcommand{\thetable}{S\arabic{table}}
\setcounter{section}{0}
\setcounter{subsection}{0}
\setcounter{equation}{0}
\setcounter{figure}{0}
\setcounter{table}{0}

\section{Numerical simulations for additional models}

This section collects numerical checks of the main-text results across several one-dimensional non-Hermitian critical free-fermion models.
We first verify the additivity formula [main-text Eq.~(15)] for band structures that combine imaginary Dirac points, single-band Fermi points, and an exceptional point.
We then confirm that the logarithmic coefficient halves under open boundary conditions in a model free of the non-Hermitian skin effect.
Next, we treat an imaginary Dirac point displaced from $\operatorname{Im}E=0$ by a tunable occupation cut.
Finally, we apply the same classification to non-Hermitian ground states, defined by crossings of $\operatorname{Re}E(k)=0$ rather than $\operatorname{Im}E(k)=0$.
In every case, the analytic coefficients follow from main-text Eqs.~(14) and (15), and the lattice fits reproduce them.

\subsection{Additivity and exceptional points}

We further consider three models.

\begin{enumerate}
\item The Su--Schrieffer--Heeger (SSH) model in main-text Eq.~(1), with
\begin{equation}
    (t_1,t_2,g_1,g_2)=(1,2,0.5,0).
\end{equation}

\item A three-orbital model with an additional orbital $s$ in each unit cell:
\begin{align}
\hat H_{\rm mix}=\sum_j\Big[
&t_1e^{g_1}c_{j,b}^\dagger c_{j,a}
+t_1e^{-g_1}c_{j,a}^\dagger c_{j,b}
+t_2e^{g_2}c_{j+1,a}^\dagger c_{j,b}
+t_2e^{-g_2}c_{j,b}^\dagger c_{j+1,a}
\nonumber\\
&+\lambda(c_{j,a}^\dagger c_{j,s}-c_{j+1,a}^\dagger c_{j,s}
+c_{j,s}^\dagger c_{j,a}-c_{j,s}^\dagger c_{j+1,a})
+\ii\gamma c_{j,s}^\dagger c_{j,s}
+t_c(c_{j+1,s}^\dagger c_{j,s}-c_{j,s}^\dagger c_{j+1,s})
\Big].
\end{align}
Under periodic boundary conditions, this gives, in the basis $(a,b,s)$,
\[
H_{\rm mix}(k)=
\begin{pmatrix}
0 & h_+(k) & \lambda(1-e^{-\ii k})\\
h_-(k) & 0 & 0\\
\lambda(1-e^{\ii k}) & 0 & \ii\gamma-2\ii t_c\sin k
\end{pmatrix},
\]
where $h_\pm(k)$ are defined as in main-text Eq.~(4).
We use
\begin{equation}
(t_1,t_2,g_1,g_2)=(3.6,2,1,0),
\qquad
\gamma=0.6,\qquad
t_c=1,\qquad
\lambda=0.6 .
\end{equation}

\item The SSH model with
\begin{equation}
    (t_1,t_2,g_1,g_2)=(2e^{-1/2},2,0.5,0).
\end{equation}
\end{enumerate}

\begin{figure}[!ht]
\centering
\includegraphics[width=0.8\linewidth]{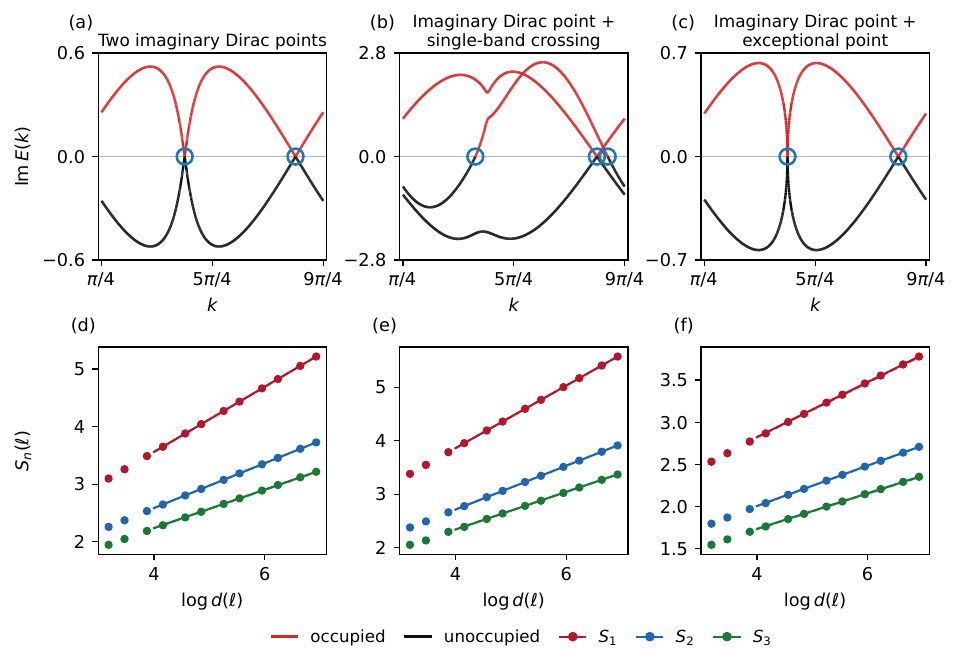}
\caption{Additivity checks for three clean critical structures.
The first row shows $\operatorname{Im}E(k)$, with occupied branches in red, unoccupied branches in black, and circles marking the momenta where $\operatorname{Im}E(k)=0$.
The second row shows the R\'enyi entropies $S_1$, $S_2$, and $S_3$ as functions of $\ln d_\ell$.
Solid lines are numerical fits, and dashed lines show the analytic predictions from the main-text additivity formula.
The fitted and analytic coefficients are listed in Table~\ref{tab:sm-additivity-examples}.}
\label{fig:sm-additivity-examples}
\end{figure}

Figure~\ref{fig:sm-additivity-examples}(a)--(c) shows the corresponding $\operatorname{Im}E(k)$ bands.
Red and black denote occupied and unoccupied branches, respectively, and the circles mark the momenta where $\operatorname{Im}E(k)=0$.
\begin{enumerate}
\item For the first model, $\operatorname{Im}E_\pm(k)=0$ at $k=0$ and $k=\pi$, corresponding to two imaginary Dirac points.
Using the overlap-angle formula in main-text Eq.~(9), we find $\theta(k=0)=0.167\ldots$ and $\theta(k=\pi)=0.640\ldots$.
The two contributions $\zeta_n(\theta_i)$, evaluated from main-text Eq.~(14), add according to main-text Eq.~(15); the resulting theoretical values are listed in Table~\ref{tab:sm-additivity-examples}.
\item For the second model, one band satisfies $\operatorname{Im}E(k)=0$ at $k=k_0,\pi-k_0$, where $k_0=\arcsin[\gamma/(2t_c)]=0.3046\ldots$, while two bands cross at $k=0$.
This gives one imaginary Dirac point and two single-band Fermi points.
The imaginary Dirac point has $\theta_{\rm mix}=0.594\ldots$, and the total coefficient is $\zeta_n(\theta_{\rm mix})+(1+1/n)/6$.
\item For the third model, two bands cross at $k=0$ and $k=\pi$.
At $k=0$ the crossing is an imaginary Dirac point with $\theta=0.189\ldots$, while at $k=\pi$
\begin{equation}
    H_{\rm SSH}(\pi)
    =
    \begin{pmatrix}
        0 & 2e^{-1}-2\\
        0 & 0
    \end{pmatrix}.
\end{equation}
The $k=\pi$ exceptional point corresponds to $\theta=\pi/2$ and contributes zero, so the total coefficient is $\zeta_n(0.189\ldots)$.
\end{enumerate}

For a periodic chain, we fit to the main-text scaling form using $L=32768$, the boundary condition $c_{j+L,\alpha}=e^{\ii\phi}c_{j,\alpha}$ with $\phi=0.37$, and the fit window $\ell\ge512$.
The twist shifts the allowed momenta away from the zeros of $\operatorname{Im}E(k)$, removing finite-size ambiguity in the occupation projector.
The second row of Fig.~\ref{fig:sm-additivity-examples} shows the numerical $S_1$, $S_2$, and $S_3$ data.
The fitted lines and analytic predictions are drawn only for $\ln d_\ell>4$.

\begin{table}[!ht]
\centering
\begingroup
\setlength{\tabcolsep}{4pt}
\renewcommand{\arraystretch}{1.08}
\begin{tabular}{@{}p{0.46\linewidth}@{\hspace{1.2em}}llll@{}}
\toprule
case & value & $n=1$ & $n=2$ & $n=3$ \\
\midrule
two imaginary Dirac points & fit & 0.565 & 0.389 & 0.333 \\
& theory & $0.5650\ldots$ & $0.3886\ldots$ & $0.3317\ldots$ \\
\addlinespace
imaginary Dirac point + single-band Fermi points & fit & 0.584 & 0.409 & 0.351 \\
& theory & $0.5841\ldots$ & $0.4088\ldots$ & $0.3520\ldots$ \\
\addlinespace
imaginary Dirac point + exceptional point & fit & 0.326 & 0.239 & 0.211 \\
& theory & $0.3245\ldots$ & $0.2389\ldots$ & $0.2097\ldots$ \\
\bottomrule
\end{tabular}
\endgroup
\caption{Comparison between the numerical slopes in Fig.~\ref{fig:sm-additivity-examples} and the analytic coefficients from the main-text additivity formula.
The larger relative deviations at higher R\'enyi index come from stronger finite-size corrections.}
\label{tab:sm-additivity-examples}
\end{table}

Table~\ref{tab:sm-additivity-examples} shows excellent agreement between the numerical fits and the theoretical coefficients.

\subsection{Anomalous entanglement scaling under boundary conditions}

\begin{figure}[!ht]
\centering
\includegraphics[width=0.98\linewidth]{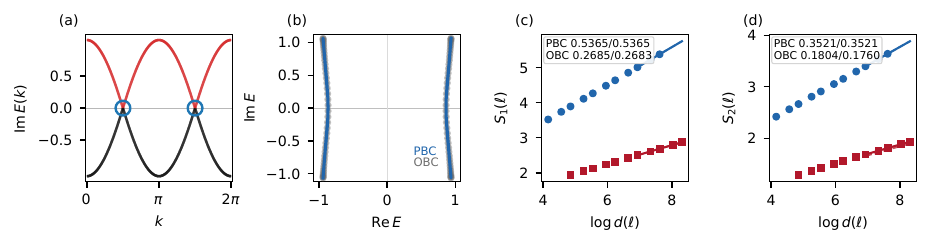}
\caption{Boundary-condition check in a reciprocal model without the non-Hermitian skin effect.
(a) Periodic-boundary imaginary-energy band; red and black denote occupied and unoccupied branches, and circles mark the imaginary Dirac points.
(b) Complex spectrum under periodic boundary conditions (PBC) and open boundary conditions (OBC); the OBC points are shown at a smaller size $L=128$ for readability, while the entropy data use $L=32768$.
(c,d) PBC and OBC edge-interval scaling for $S_1$ and $S_2$; solid lines are numerical fits, dashed lines use the analytic slopes $2\zeta_n(\theta)$ and $\zeta_n(\theta)$, and the slope values are written in each panel.}
\label{fig:sm-boundary-condition}
\end{figure}

We next verify that the logarithmic coefficient is halved under open boundary conditions in a reciprocal model without the non-Hermitian skin effect.
We use the two-orbital real-space Hamiltonian
\begin{align}
    \hat H
    =
    &\sum_{j=1}^{L}\Big[
    \Delta(c_{j,a}^\dagger c_{j,a}-c_{j,b}^\dagger c_{j,b})
    +\ii\gamma(c_{j,a}^\dagger c_{j,b}+c_{j,b}^\dagger c_{j,a})
    \Big]
    \nonumber\\
    &+\sum_{j=1}^{L-1}\Big[
    \frac{\ii v}{2}(c_{j+1,a}^\dagger c_{j,a}+c_{j,a}^\dagger c_{j+1,a})
    -\frac{\ii v}{2}(c_{j+1,b}^\dagger c_{j,b}+c_{j,b}^\dagger c_{j+1,b})
    \Big].
\end{align}
With periodic boundary conditions, the same hopping pattern gives
\begin{equation}
    H(k)=(\Delta+\ii v\cos k)\sigma_z+\ii\gamma\sigma_x ,
\end{equation}
with
\begin{equation}
    \Delta=1,\qquad \gamma=0.5,\qquad v=1 .
\end{equation}
The periodic-boundary imaginary-energy zeros are at $k=\pi/2,3\pi/2$, as shown in Fig.~\ref{fig:sm-boundary-condition}(a).
The complex spectra under periodic and open boundary conditions lie on the same curve in Fig.~\ref{fig:sm-boundary-condition}(b), showing that this model has no spectral collapse associated with the non-Hermitian skin effect.
The two imaginary Dirac points are related by $k\to-k$ and therefore share the same overlap angle,
\begin{equation}
    |\sin\theta|=0.5,\qquad
    \cos\theta=\sqrt{3}/2 ,
\end{equation}
so each contributes an equal $\zeta_n(\theta)$.
For an interval in a periodic chain, the expected coefficient is therefore $2\zeta_n(\theta)$, while for an edge interval in an open chain it is $\zeta_n(\theta)$.

For the open-boundary numerics, we use the standing-wave representation rather than diagonalizing the full $2L\times2L$ matrix.
The allowed momenta are $q_m=m\pi/(L+1)$.
For each $q_m$, we diagonalize the $2\times2$ matrix $H(q_m)$, form the occupied projector $P_+(q_m)$, and assemble the edge correlation matrix
\begin{equation}
    C_{ij}
    =
    \frac{2}{L+1}
    \sum_{m=1}^{L}
    \sin(q_m i)\sin(q_m j)P_+(q_m).
\end{equation}
The R\'enyi entropies are then obtained from the eigenvalues of this restricted correlation matrix.
The data in Fig.~\ref{fig:sm-boundary-condition}(c,d) use $L=32768$ for both boundary conditions and fit the large-distance window $\ell\ge1024$.
The agreement is quantitative for $S_1$ and remains good for $S_2$; the slightly larger open-boundary deviation at $n=2$ is due to the stronger finite-size corrections at higher R\'enyi index.

\FloatBarrier

\subsection{Off-zero imaginary Dirac point}

We next consider a nearest-neighbor two-orbital model whose occupation boundary need not be $\operatorname{Im}E=0$.
The real-space Hamiltonian is
\begin{align}
\hat H=\sum_j\Big[
    &\Delta(c_{j,a}^\dagger c_{j,a}-c_{j,b}^\dagger c_{j,b})
    +\ii\gamma(c_{j,a}^\dagger c_{j,b}+c_{j,b}^\dagger c_{j,a})
    +\frac{\ii v}{2}(c_{j+1,a}^\dagger c_{j,a}+c_{j,a}^\dagger c_{j+1,a})
    -\frac{\ii v}{2}(c_{j+1,b}^\dagger c_{j,b}+c_{j,b}^\dagger c_{j+1,b})
    \nonumber\\
    &+\frac{w}{2}\sum_{\alpha=a,b}
    (c_{j,\alpha}^\dagger c_{j+1,\alpha}
    -c_{j+1,\alpha}^\dagger c_{j,\alpha})
    \Big].
\end{align}
Under periodic boundary conditions,
\begin{equation}
    H(k)=\ii w\sin k\,I_2+(\Delta+\ii v\cos k)\sigma_z+\ii\gamma\sigma_x .
\end{equation}
We use
\begin{equation}
    \Delta=1,\qquad \gamma=0.5,\qquad v=1,\qquad w=0.4 .
\end{equation}
At fixed filling, we occupy modes with $\operatorname{Im}E>\mu_I$.

\begin{figure}[!ht]
\centering
\includegraphics[width=0.98\linewidth]{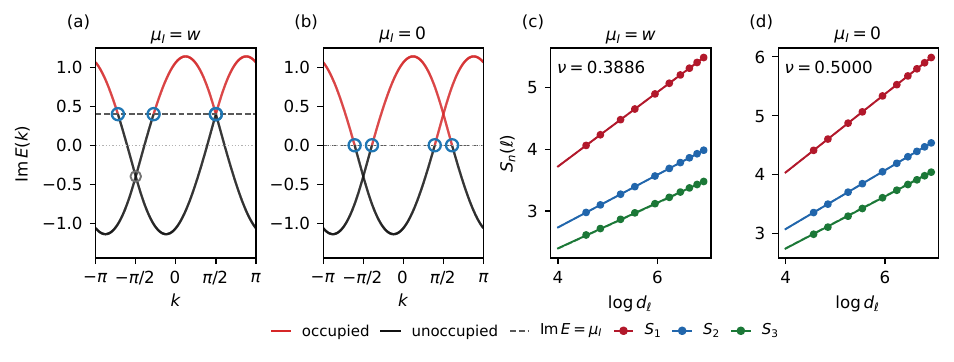}
\caption{Off-zero imaginary Dirac point and the corresponding entropy scaling.
(a,b) Imaginary-energy bands for the cuts $\mu_I=w$ and $\mu_I=0$; the gray circle in (a) marks the second imaginary Dirac point, which does not lie on the $\mu_I=w$ occupation boundary.
(c,d) R\'enyi entropies for the two cuts; solid lines are numerical fits, and dashed lines show the analytic predictions.
The fitted and analytic coefficients are listed in Table~\ref{tab:sm-off-zero-idp}.}
\label{fig:sm-off-zero-idp}
\end{figure}

Figure~\ref{fig:sm-off-zero-idp}(a,b) shows the imaginary-energy bands for two choices of the occupation cut.
Red and black denote occupied and unoccupied branches, respectively, and the circles mark the crossings on the chosen occupation boundary.
\begin{enumerate}
\item For $\mu_I=w$, the crossing at $k=\pi/2$ lies on the occupation boundary:
\begin{equation}
    E_\pm(\pi/2)=\ii w\pm\sqrt{\Delta^2-\gamma^2}
    =0.4\ii\pm\frac{\sqrt{3}}{2}.
\end{equation}
This is an off-zero imaginary Dirac point with
\begin{equation}
    |\sin\theta|=\frac{\gamma}{\Delta}=0.5,\qquad
    \cos\theta=\frac{\sqrt{3}}{2}.
\end{equation}
The same cut also has two single-band Fermi points at $k=-0.722861\pi$ and $k=-0.277139\pi$, and the filling is $\nu=0.388565\ldots$.
The theoretical coefficient is
\begin{equation}
    \zeta_n(\theta)+\frac{1+1/n}{6}.
\end{equation}
\item For $\mu_I=0$, the cut has four single-band Fermi points and no imaginary Dirac point.
The filling is $\nu=1/2$, and the theoretical coefficient is
\begin{equation}
    \frac{1+1/n}{3}.
\end{equation}
\end{enumerate}

For both cuts, we use $L=65536$, twisted momenta $k_m=2\pi(m+0.37)/L$, and fit the periodic-boundary entropy to
\begin{equation}
    S_n(\ell)=a_n+\zeta_n\ln\left[\frac{L}{\pi}
    \sin\frac{\pi\ell}{L}\right]
\end{equation}
over $\ell\ge512$.
Figure~\ref{fig:sm-off-zero-idp}(c,d) shows the numerical data, the fitted lines, and the analytic predictions.

\FloatBarrier

\begin{table}[!ht]
\centering
\begingroup
\setlength{\tabcolsep}{6pt}
\renewcommand{\arraystretch}{1.08}
\begin{tabular}{lllll}
\toprule
cut & value & \(n=1\) & \(n=2\) & \(n=3\) \\
\midrule
\(\mu_I=w\) & fit & 0.602 & 0.426 & 0.369 \\
 & theory & \(0.6016\ldots\) & \(0.4260\ldots\) & \(0.3680\ldots\) \\
\addlinespace
\(\mu_I=0\) & fit & 0.667 & 0.500 & 0.443 \\
 & theory & \(0.6667\ldots\) & \(0.5000\ldots\) & \(0.4444\ldots\) \\
\bottomrule
\end{tabular}
\endgroup
\caption{Comparison between the numerical slopes in Fig.~\ref{fig:sm-off-zero-idp} and the analytic coefficients from the additivity formula.}
\label{tab:sm-off-zero-idp}
\end{table}

Table~\ref{tab:sm-off-zero-idp} shows excellent agreement between the numerical fits and the theoretical coefficients.
The higher R\'enyi entropies show slightly larger finite-size corrections.

\FloatBarrier

\subsection{Ground states}

So far, we have filled the bands by $\operatorname{Im}E$, the occupation selected by the non-Hermitian time evolution.
One is often equally interested in the ground state, the non-Hermitian counterpart of the familiar Hermitian ground state.
Just as the Hermitian ground state fills every negative-energy single-particle mode, we define the non-Hermitian ground state as the many-body right eigenstate with the smallest $\operatorname{Re}E$, obtained by occupying every single-particle right eigenmode with $\operatorname{Re}E<0$.
Following the main text, its logarithmic entanglement is then controlled by the crossings of the occupation boundary $\operatorname{Re}E(k)=0$ rather than $\operatorname{Im}E(k)=0$, with a two-band crossing contributing $\zeta_n(\theta)$ [main-text Eq.~(14)] through the overlap angle $\theta$ of the right spinors there [main-text Eq.~(9)].

We illustrate this with two ground-state fillings:
\begin{enumerate}
\item The SSH model [main-text Eq.~(1)] with
\begin{equation}
    (t_1,t_2,g_1,g_2)=(3.6,2,1,0).
\end{equation}

\item The same model with
\begin{equation}
    (t_1,t_2,g_1,g_2)=(1,2,0.5,0).
\end{equation}
\end{enumerate}

\begin{figure}[!ht]
\centering
\includegraphics[width=0.98\linewidth]{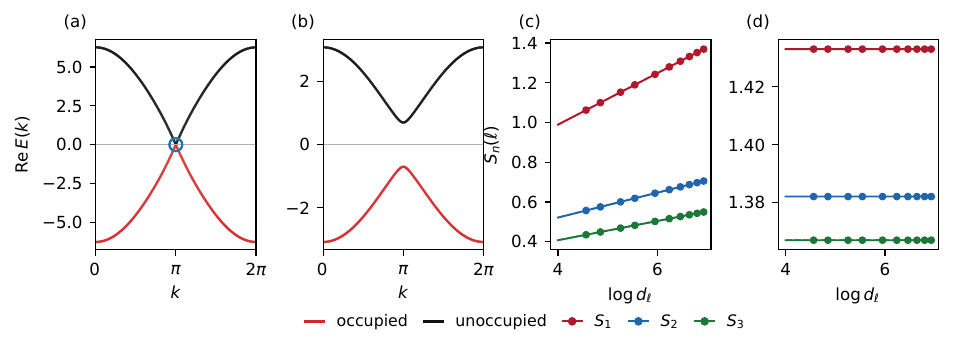}
\caption{Real-line ground-state fillings.
(a,b) Real-energy bands for the weakened-discontinuity and gapped cases; red and black denote occupied and unoccupied branches, respectively, and the circle in (a) marks the real-line Fermi point.
(c,d) R\'enyi entropies $S_1$, $S_2$, and $S_3$; solid lines are numerical fits, and dashed lines show the analytic predictions.
The fitted and analytic coefficients are listed in Table~\ref{tab:sm-ground-state}.}
\label{fig:sm-ground-state}
\end{figure}

Figure~\ref{fig:sm-ground-state}(a,b) shows the corresponding $\operatorname{Re}E(k)$ bands.
Red and black denote occupied and unoccupied branches, respectively, and the circle in panel (a) marks the momentum where $\operatorname{Re}E(k)=0$.
\begin{enumerate}
\item For the first model, the occupation has a weakened discontinuity at $k=\pi$, where the two bands cross on the boundary $\operatorname{Re}E=0$.
The overlap angle of the right spinors there is $\theta_{\rm gs}$ with $|\sin\theta_{\rm gs}|=0.840\ldots$, so the coefficients are $\zeta_n(\theta_{\rm gs})$.
\item For the second model, the spectrum is gapped across $\operatorname{Re}E=0$.
There is no Fermi point, and the logarithmic coefficient vanishes.
\end{enumerate}

For both cases, we use $L=65536$, twisted momenta $k_m=2\pi(m+0.37)/L$, and fit the periodic-boundary entropy to
\begin{equation}
    S_n(\ell)=a_n+\zeta_n\ln\left[\frac{L}{\pi}
    \sin\frac{\pi\ell}{L}\right]
\end{equation}
over $\ell\ge256$.
Figure~\ref{fig:sm-ground-state}(c,d) shows the numerical data, the fitted lines, and the analytic predictions.

\FloatBarrier

\begin{table}[!ht]
\centering
\begingroup
\setlength{\tabcolsep}{6pt}
\renewcommand{\arraystretch}{1.08}
\begin{tabular}{@{}lllll@{}}
\toprule
case & value & \(n=1\) & \(n=2\) & \(n=3\) \\
\midrule
partial jump & fit & 0.130 & 0.063 & 0.048 \\
 & theory & \(0.1297\ldots\) & \(0.0627\ldots\) & \(0.0484\ldots\) \\
\addlinespace
gapped & fit & 0.000 & 0.000 & 0.000 \\
 & theory & 0 & 0 & 0 \\
\bottomrule
\end{tabular}
\endgroup
\caption{Comparison between the numerical slopes in Fig.~\ref{fig:sm-ground-state} and the analytic coefficients.}
\label{tab:sm-ground-state}
\end{table}

\FloatBarrier

Table~\ref{tab:sm-ground-state} shows excellent agreement between the numerical fits and the theoretical coefficients.

\FloatBarrier

\section{Lattice scattering from a local onsite impurity}
\label{sec:sm-scatt-scattering}
In the main text, the robustness of the logarithmic scaling to weak onsite disorder was traced to the absence of impurity backscattering.
We substantiate this here by solving the single-impurity scattering problem directly on the lattice and showing that a local onsite impurity produces no reflected propagating wave.
Consider a scalar onsite impurity at cell $0$,
\begin{equation}
V_{\rm imp}=u\,(c_{0,a}^\dagger c_{0,a}+c_{0,b}^\dagger c_{0,b}).
\end{equation}
Let $a_j$ and $b_j$ denote the single-particle wave-function amplitudes on sublattices $A$ and $B$ in unit cell $j$, respectively.
For $j\ll0$ the incoming wave is a periodic-boundary Bloch state at complex energy $E$,
\begin{equation}
\begin{pmatrix}
a_j^{\rm in}\\
b_j^{\rm in}
\end{pmatrix}
=e^{\ii k j}u_k,\qquad k\in\mathbb R ,
\label{eq:sm-scatt-incoming-bloch}
\end{equation}
where $u_k$ is the two-component Bloch spinor.
We propagate the amplitudes with the transfer vector $\Phi_j=(a_j,b_{j-1})^T$, chosen so that a single step advances $(a_j,b_{j-1})$ to $(a_{j+1},b_j)$ and keeps both onsite terms of a unit cell within one transfer matrix.
Away from the impurity, the clean lattice equations are
\begin{align*}
E a_j&=t_1e^{-g_1}b_j+t_2e^{g_2}b_{j-1},\\
E b_j&=t_1e^{g_1}a_j+t_2e^{-g_2}a_{j+1}.
\end{align*}
Solving these equations for $b_j$ and $a_{j+1}$ gives
\begin{align*}
b_j&=
\frac{E e^{g_1}}{t_1}a_j
-\frac{t_2e^{g_1+g_2}}{t_1}b_{j-1},
\\
a_{j+1}
&=
\frac{e^{g_1+g_2}}{t_2}
\left(\frac{E^2}{t_1}-t_1\right)a_j
-\frac{E e^{g_1+2g_2}}{t_1}b_{j-1}.
\end{align*}
Thus
\begin{equation}
\Phi_{j+1}=T_0(E)\Phi_j,
\end{equation}
with
\begingroup
\setlength{\arraycolsep}{3pt}
\begin{equation}
T_0(E)=
\begin{pmatrix}
\frac{e^{g_1+g_2}}{t_2}\left(\frac{E^2}{t_1}-t_1\right)
&
-\frac{E e^{g_1+2g_2}}{t_1}
\\[2mm]
\frac{E e^{g_1}}{t_1}
&
-\frac{t_2e^{g_1+g_2}}{t_1}
\end{pmatrix}.
\label{eq:sm-scatt-clean-transfer}
\end{equation}
\endgroup
At the impurity cell the onsite term replaces $E$ by $E-u$, so crossing the impurity gives
\begin{equation}
\Phi_1=T_u(E)\Phi_0,
\end{equation}
where $T_u(E)$ is obtained from $T_0(E)$ by $E\to E-u$:
\begingroup
\setlength{\arraycolsep}{3pt}
\begin{equation}
T_u(E)=
\begin{pmatrix}
\frac{e^{g_1+g_2}}{t_2}\left(\frac{(E-u)^2}{t_1}-t_1\right)
&
-\frac{(E-u)e^{g_1+2g_2}}{t_1}
\\[2mm]
\frac{(E-u)e^{g_1}}{t_1}
&
-\frac{t_2e^{g_1+g_2}}{t_1}
\end{pmatrix}.
\label{eq:sm-scatt-impurity-transfer}
\end{equation}
\endgroup
Each transfer eigenvalue $\lambda$ is associated with the eigenmode
\begin{equation}
T_0(E)
\begin{pmatrix}
1\\
\eta_\lambda
\end{pmatrix}
=
\lambda
\begin{pmatrix}
1\\
\eta_\lambda
\end{pmatrix},
\qquad
\eta_\lambda=
\frac{E e^{g_1}}{t_1\lambda+t_2e^{g_1+g_2}} ,
\label{eq:sm-scatt-transfer-eigenvector}
\end{equation}
The two eigenvalues relevant here are $\lambda=e^{\ii k}$ and $\lambda=w^{-1}$.
The physical Bloch spinor of the incoming eigenvalue is $u_k=(1,e^{\ii k}\eta_{e^{\ii k}})^T$, so in transfer variables the incoming state Eq.~\eqref{eq:sm-scatt-incoming-bloch} becomes
\begin{equation}
\Phi_j^{\rm in}
=e^{\ii k j}
\begin{pmatrix}
1\\
\eta_{e^{\ii k}}
\end{pmatrix},
\qquad k\in\mathbb R .
\label{eq:sm-scatt-incoming-transfer}
\end{equation}
At the same energy, the second transfer eigenvalue is fixed by $\det T_0(E)=e^{2(g_1+g_2)}$; we denote it $w^{-1}$, with
\begin{equation}
w=e^{-2(g_1+g_2)}e^{\ii k}.
\label{eq:sm-scatt-partner-root}
\end{equation}
For real $k$ and $g_1+g_2>0$, $|w|=e^{-2(g_1+g_2)}<1$, so $|w^{-1}|>1$ and this eigenvalue is evanescent rather than a second propagating Bloch wave.
The partner mode $w^{-j}(1,\eta_{w^{-1}})^T$ decays as $j\to-\infty$ and grows as $j\to+\infty$, so it can enter only as a localized tail on the incoming side, never as an outgoing wave on the right.
At each complex energy, the chain therefore supports a single propagating channel---the lattice counterpart of the main-text statement that each complex $E$ is carried by one periodic-boundary state.

The bounded left-incident scattering state is therefore
\begin{equation}
\Phi_j=
\begin{cases}
e^{\ii k j}
\begin{pmatrix}1\\ \eta_{e^{\ii k}}\end{pmatrix}
+r\,w^{-j}
\begin{pmatrix}1\\ \eta_{w^{-1}}\end{pmatrix},
& j\le0,
\\[2mm]
t\,e^{\ii k j}
\begin{pmatrix}1\\ \eta_{e^{\ii k}}\end{pmatrix},
& j\ge1 .
\end{cases}
\label{eq:sm-scatt-state}
\end{equation}
The incoming and transmitted waves occupy the same propagating Bloch channel, with amplitude rescaled from $1$ to $t$, while $r$ multiplies only the evanescent tail rather than a reflected Bloch wave.
A local onsite impurity therefore opens no counter-propagating channel, and backscattering is absent---the lattice origin of the robustness of the logarithmic scaling to weak onsite disorder.

For completeness, the evanescent-tail amplitude and the transmission amplitude follow from the matching condition
\begin{equation}
T_u(E)\left[
\begin{pmatrix}1\\ \eta_{e^{\ii k}}\end{pmatrix}
+r
\begin{pmatrix}1\\ \eta_{w^{-1}}\end{pmatrix}
\right]
=t\,e^{\ii k}
\begin{pmatrix}1\\ \eta_{e^{\ii k}}\end{pmatrix}.
\label{eq:sm-scatt-scattering-match}
\end{equation}
With the shorthand
\begin{equation}
\begin{aligned}
A_\lambda=
-\frac{u}{t_1(\eta_{w^{-1}}-\eta_{e^{\ii k}})}
\left[
e^{g_1}
+\eta_{e^{\ii k}}
\left\{
\frac{e^{g_1+g_2}}{t_2}(u-2E)
+e^{g_1+2g_2}\eta_\lambda
\right\}
\right],
\qquad \lambda=e^{\ii k},w^{-1} .
\end{aligned}
\label{eq:sm-scatt-A-lambda}
\end{equation}
the solution is
\begin{equation}
r=-\frac{w A_{e^{\ii k}}}{1+w A_{w^{-1}}},
\qquad
t=\frac{1}{1+w A_{w^{-1}}}.
\label{eq:sm-scatt-rt-lattice}
\end{equation}
In the clean limit $u=0$, $A_{e^{\ii k}}=A_{w^{-1}}=0$ and $(r,t)=(0,1)$.